\documentclass[conference]{IEEEtran}
\IEEEoverridecommandlockouts
\usepackage{cite}
\usepackage{amsmath,amssymb,amsfonts}
\usepackage{algorithmic}
\usepackage{graphicx}
\usepackage{textcomp}
\usepackage{xcolor}
\def\BibTeX{{\rm B\kern-.05em{\sc i\kern-.025em b}\kern-.08em
    T\kern-.1667em\lower.7ex\hbox{E}\kern-.125emX}}

\begin{document}

\title{A Dynamic Phasor Framework for Analysis of Grid-Forming Converter Connected to Series-Compensated Line\\
\thanks{Financial support from NSF Grant Award ECCS 2317272 is gratefully acknowledged.}
}

\author{\IEEEauthorblockN{Fiaz Hossain and Nilanjan Ray Chaudhuri}
\IEEEauthorblockA{\textit{School of Electrical Engineering and Computer
Science}, The Pennsylvania State University, University Park, PA, USA \\
emails: fbh5142@psu.edu, nuc88@psu.edu
}}

\maketitle 

\begin{abstract}
A dynamic phasor (DP) framework for time-domain and frequency-domain analyses of grid-forming converters (GFCs) connected to series-compensated transmission lines is proposed. The proposed framework can capture the behavior of GFCs subjected to unbalanced short circuit faults in presence of different current limiting strategies. Moreover, the linearizability and time invariance of this framework allows us to perform eigen decomposition, which is a powerful tool for root-cause analysis and control design. We show that a certain degree of series compensation may result in poorly-damped oscillations in presence of the grid-forming converter. A participation factor analysis using the DP model reveals that the point of interconnection voltage angle is dominant in this mode. Eigenvalue sensitivity analysis of controller parameters shows that reducing the power-frequency droop coefficient is  most effective in stabilizing the poorly-damped mode. Detailed validation with electromagnetic transient model demonstrates the accuracy of the proposed framework.  
\end{abstract}

\begin{IEEEkeywords}
Dynamic Phasor, Grid Forming Converter
\end{IEEEkeywords}

\section{Introduction}
Traditional planning models of power systems are based upon the positive sequence fundamental frequency phasor framework. Although this framework has served well for many decades in adequately capturing the dynamics of bulk grids with synchronous generators (SGs), it is being challenged with the advent of inverter-based resources (IBRs). Today, commercial simulation softwares have included generic dynamic models of IBRs in their library and does have the provision of including user-defined models. However, these models cannot simulate unbalanced short circuit faults.

Unlike SG-dominated grids, where only balanced three-phase faults are simulated since they represent the worst-case scenario from transient stability standpoint, this may not be adequate for IBR-dominated grids. Unbalanced faults are more common than the balanced ones, and IBR response to such faults could lead to some major issues in the grid. This has recently been observed in the response of the IBRs where partial reduction of power output took place \cite{NERC_PV_CAISO}.

To address these challenges, North American Electric Reliability Corporation (NERC) has developed guidelines for electromagnetic transient (EMT) modeling of IBRs connected to bulk power systems \cite{NERC_EMT}. EMT modeling framework has the ability to capture the fastest transients in the system and study unbalanced short circuit faults. However, it has certain limitations including: (a) scalability issues due to heavy computational burden and (b) unsuitability in deriving linear time invariant models from nonlinear EMT models in commercial packages that allow frequency domain analysis.




As the number of IBRs grow in the system, emerging oscillation issues including subsynchronous oscillations (SSOs) are being observed \cite{IBRSSOTF}. This demands modeling frameworks that have the ability to use well-established linear system theoretic tools to help determine the root cause of such phenomena and remedy them using holistic system-level control design. 

As a \textit{middle ground} that lies between traditional phasor and high resolution EMT frameworks, dynamic phasor (DP)-based modeling framework has been proposed in literature \cite{Verghese-91-DP,Stankovic-DPfaults} that has the following attributes -\\
(a) It is capable of simulating \textit{unbalanced systems} including unbalanced short-circuit faults.~
(b) The framework is \textit{expected to be scalable} as it takes advantage of variable time-step numerical solvers.~
(c) The framework is \textit{linearizable} \textit{and time invariant} and therefore is suitable for eigen decomposition-based frequency-domain analysis and control design.~
(d) It has a \textit{tunable degree of complexity-vs-accuracy trade-off} based on the order of DPs considered in the model.

\textit{We believe that the DP and the EMT frameworks are complementary to each other and each has its own place in the analysis of grids with high IBR penetration.} With that spirit, in this paper we present a DP-based modeling framework for grid-forming converters (GFCs) connected to series-compensated lines. 

In \cite{GFC_DP}, a GFC model in DP framework was presented. However, the paper lacks details on what order of DPs were considered and it did not model current limiting features of the GFC. It is not clear if the model has capability to simulate unbalanced fault with nonzero fault impedance. Also, no frequency-domain analysis was presented, since the focus was only on time-domain analysis.

The contributions of our paper are the following - \\
(1) We present a DP-based model of the GFC that can be interfaced with an unbalanced grid model including non-zero fault impedance. Our GFC model is built in the rotating $dq$ frame of its controller while the network is modeled in $pnz$ frame. \\
(2) We have proposed an approach to include current limiting features in the DP-based GFC model. \\
(3) To the best of our knowledge, this is the first work that studies a GFC connected to a series-compensated line. To this end, we have shown that beyond a certain level of compensation, poorly-damped modes may arise. \\
(4) We have performed linearization of the DP-based model to determine the root cause of the mode and subsequently presented a holistic approach to improve the damping of that mode. \\
(5) Finally, we have compared the results of DP-based model with EMT simulations to demonstrate a very close match between the two. 
\vspace{-2pt}

\section{Fundamentals of Dynamic Phasor (DP)}
 In \cite{Verghese-91-DP}, the generalized averaging theory was proposed, which expresses a near-periodic (possibly complex) time-domain waveform $x(\tau)$ in the interval $\tau \in (t - T, t]$ using a Fourier series of the form $x(\tau) = \sum_{k = -\infty}^{\infty} X_k(t) e^{jk\omega_s\tau}$, where $\omega_s = \frac{2\pi}{T}$, $k \in \mathbb{Z}$, and $X_k(t)$ are the complex Fourier coefficients that vary with time as the window of width $T$ slides over the signal. The $k$th coefficient, also called the \textit{$k$th dynamic phasor (DP)}, can be determined at time $t$ by the following \textit{averaging} operation $X_k(t) = \frac{1}{T}\int_{t-T}^{t} x(\tau) e^{-jk\omega_s\tau}d\tau = \left \langle x \right \rangle_k (t)$. In DP framework we are interested in a good approximation provided by the set $\mathcal{U}$ of dominant Fourier coefficients such that $x(\tau) \approx  \sum_{k \in \mathcal{U}} \left \langle x \right \rangle_{k}(t) e^{jk\omega_s\tau}$. Therefore, the generalized averaging-based method leads to an approximated model. From now on, we drop the time variable from DP notations for sake of simplicity. The following are some of the useful properties of DPs. \\
1. The derivative operation satisfies the following relation $    \left \langle \frac{\mathrm{d} x}{\mathrm{d} t} \right \rangle_k = \frac{\mathrm{d} \left \langle x \right \rangle_{k}}{\mathrm{d} t} + jk\omega_s\left \langle x \right \rangle_{k}$.\\
2. If $x(\tau)$ is real, then we can write $    \langle x \rangle_k=\langle x \rangle_{-k}^*$.

Note that the time-domain waveform $x(\tau)$ above can be $abc$ phase quantities, asynchronous $dq0$ frame quantities, or synchronous $DQ0$ frame quantities. Using the transformation in synchronously rotating reference frame, the relationship between $DQ0$ and $pnz$ quantities can be obtained in terms of DPs. 
\vspace{-3pt}
\small
\begin{equation}\label{eqn:DQtopnz}
    \begin{aligned}
        &\langle x_D\rangle_k=\frac{\langle x_p\rangle_{k+1}+\langle x_n\rangle_{k-1}}{\sqrt{2}};~
        \langle x_Q\rangle_k=\frac{\langle x_p\rangle_{k+1}-\langle x_n\rangle_{k-1}}{\sqrt{2}j}\\
        &\langle x_p\rangle_{k+1}=\frac{\langle x_D\rangle_{k}+j\langle x_Q\rangle_{k}}{\sqrt{2}};~~
        \langle x_n\rangle_{k-1}=\frac{\langle x_D\rangle_{k}-j\langle x_Q\rangle_{k}}{\sqrt{2}}\\
       & \langle x_z \rangle_k = \langle x_0 \rangle_k 
    \end{aligned}
\end{equation}
\normalsize
The following equation describes the relationship between synchronous $DQ$ frame quantities and asynchronous $dq$ frame quantities
\vspace{-3pt}
\small
\begin{equation}\label{eqn:dqtoDQ}
    \langle x_D \rangle_k+j\langle x_Q \rangle_k = \left(\langle x_d \rangle_k+j\langle x_q \rangle_k\right)e^{j\langle\theta_c\rangle_0}
\end{equation}
\normalsize
\vspace{-3pt}
where, $\theta_c (\tau) \approx \langle\theta_c\rangle_0,~\tau \in (t - T, t]$, i.e., higher frequency variation in $\theta_c$ shown in Fig. \ref{fig:Outer}(a) is ignored.

\section{DP-Based Modeling for Unbalanced Conditions}

\subsection{Test system and proposed modeling framework}\label{sec:TestSysDPFrmwk} 
\begin{figure}
	\vspace{-5pt}
	\centering	\includegraphics[width= 0.5\textwidth]{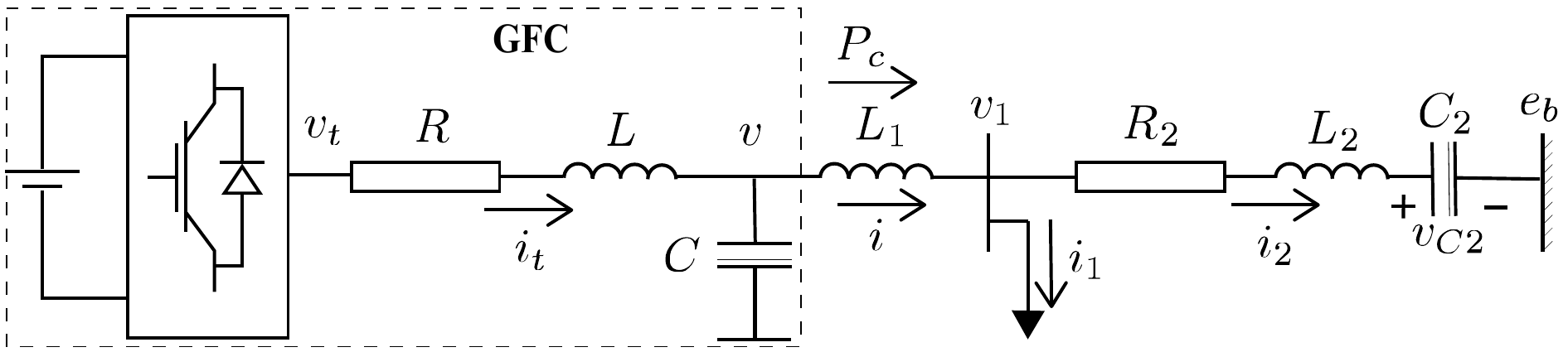}
	\caption{Single-line diagram of the simulated system. }
	\label{fig:TestSys}
	\vspace{-5pt}
\end{figure}

In this paper we focus on a test system consisting of a GFC connected to an infinite bus through a series-compensated transmission line as shown in Fig.~\ref{fig:TestSys}. A transformer interfaces the GFC and a load bus at the sending end of the line. In our model, all parameters are referred to the low voltage side of the transformer and it is represented by its leakage inductance $L_1$. The transmission line is modeled by a lumped series $R$-$L$ circuit, which is connected to a series capacitor.  

\begin{figure}
	\vspace{-5pt}
	\centering
\includegraphics[width= 0.5\textwidth]{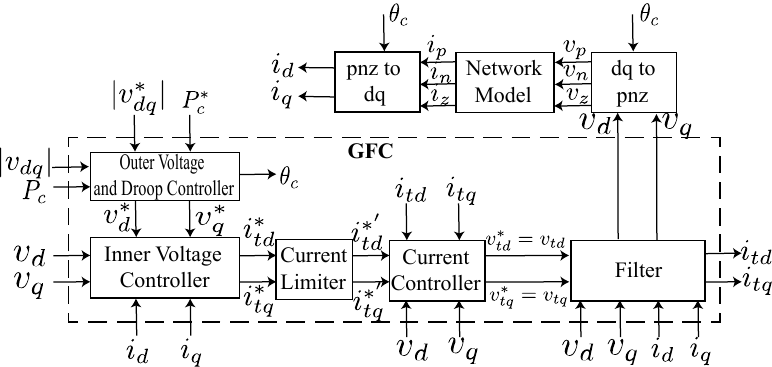}
	\caption{Proposed DP-based modeling framework showing interface between GFC and transmission network model.}
\label{fig:Framework}
	\vspace{-10pt}
\end{figure}

We propose a novel DP-based modeling framework that is capable of interfacing a GFC under \textit{conventional vector control} in rotating $dq$ frame with the rest of the system \textit{under unbalanced condition} (see Fig.~\ref{fig:Framework}), which has the following attributes.\\
1. \textit{The GFC model in $dq$-frame DPs:} Since the GFC uses conventional vector control in a rotating $dq$ frame, it is convenient to model the converter and the controller in the same $dq$ frame. This also makes it easier to adopt current limiting features in such models. The overall structure of the GFC model with input and output signals are shown in Fig.~\ref{fig:Framework}, which will be elaborated in the next subsection. Note that zero sequence is absent since there is no path for zero sequence current flow through the GFC.\\
2. \textit{The rest of the network in $pnz$-frame DPs:} The transmission line along with load and infinite bus are modeled using $pnz$-frame DPs instead of $dq0$-frame DPs. The reasons behind this are twofold: (a) the analysis of unbalanced loading and unbalanced faults can be readily accommodated using standard network parameters available in $pnz$ frame, and (b) any generic unbalanced fault (high or low resistance) can be easily modeled using the bus fault impedance matrix concept. \\
3. \textit{Interfacing GFC and network models:} This is done in two steps - first using \eqref{eqn:dqtoDQ} followed by \eqref{eqn:DQtopnz} and \textit{vice-versa}.

\subsection{GFC model with current limiting features}
\begin{figure}
	\vspace{-5pt}
	\centering
	\includegraphics[width= 0.5\textwidth]{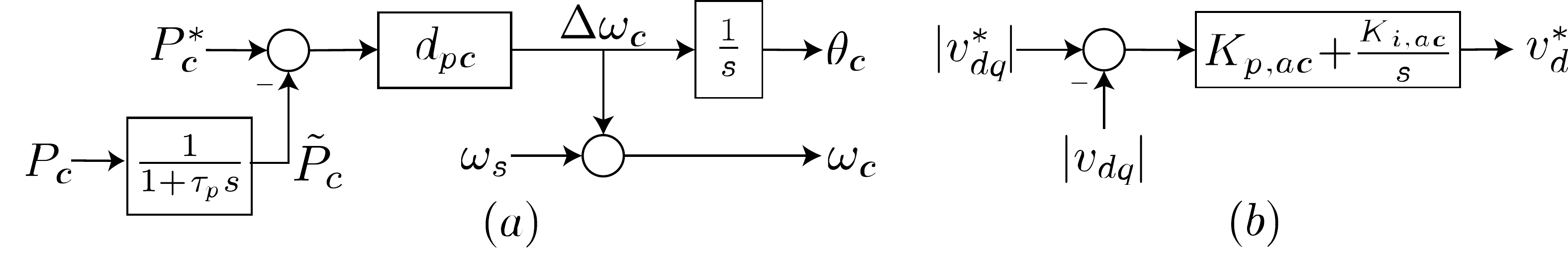}
	\caption{(a) Droop control and (b) outer voltage control loop.}
	\label{fig:Outer}
	\vspace{-5pt}
\end{figure}

\begin{figure}
	\vspace{-5pt}
	\centering
	\includegraphics[width= 0.4\textwidth]{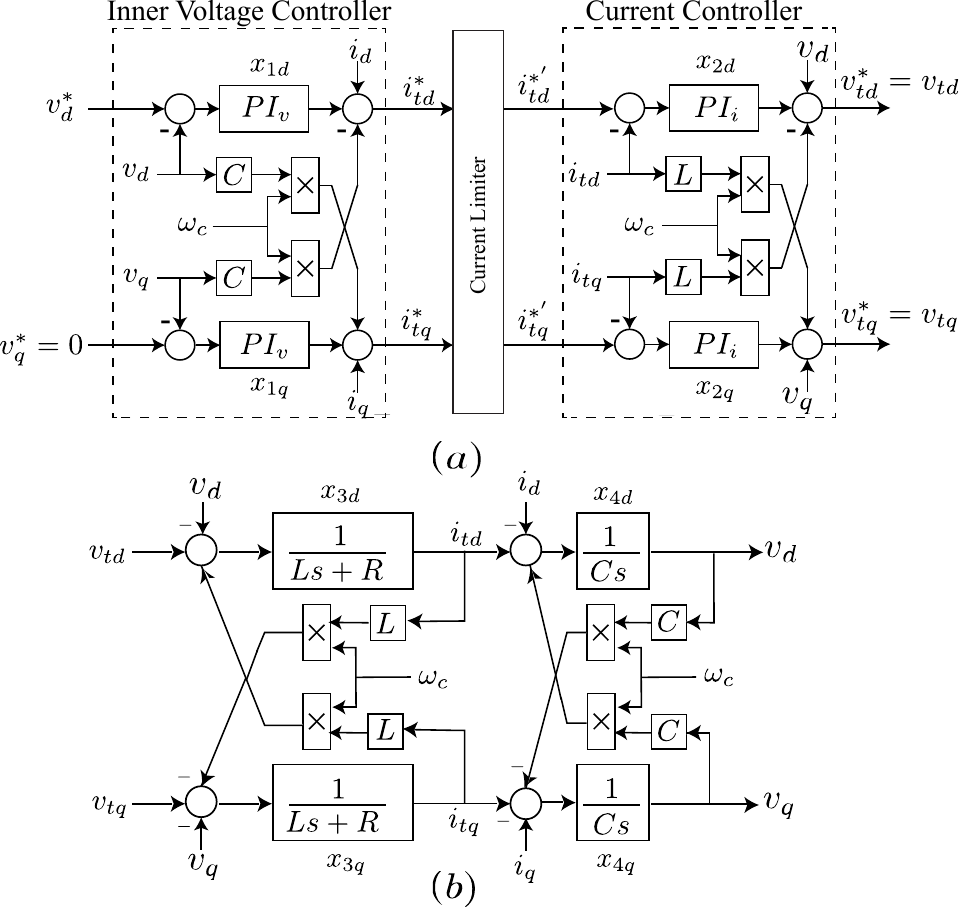}
	 \vspace{-10pt}
	\caption{(a) Inner current and voltage control loops and (b) filter.}
	\label{fig:inner}
	\vspace{-15pt}
\end{figure}

In absence of zero sequence current path, $k=0,\pm2$ are considered in the GFC $dq$-frame DP model. Throughout the paper, the following relationships hold $\langle x_{dq}\rangle_k = [\langle x_{d}\rangle_k ~\langle x_{q}\rangle_k]^T$ and $\langle x_{pnz}\rangle_k = [\langle x_{p}\rangle_k ~\langle x_{n}\rangle_k ~\langle x_{z}\rangle_k]^T$. Assuming ideal converter, the GFC model can be divided into outer voltage and droop controller as shown in Fig. \ref{fig:Outer} and inner voltage controller, current controller, and filter as shown in Fig. \ref{fig:inner}. Standard meaning is assumed for all the variables in these diagrams. DP modeling of each is described below. \\
1. \textit{Outer voltage and droop controller (Fig.~\ref{fig:Outer}):} This part of the model generates reference voltage and asynchronous $dq$ frame angle with respect to synchronous $DQ$ frame. It can be modeled in DP form using 
\small
\eqref{eqn:outer}. 
\vspace{-3pt}
\begin{equation}\label{eqn:outer}
\begin{aligned}
    &\langle \dot{v}_d^* \rangle_k=K_{i,ac}\left(|\langle v_{dq}^*\rangle_k|-|\langle v_{dq}\rangle_k|\right)\\&\langle v_d^* \rangle_k = K_{p,ac}\left(|\langle v_{dq}^*\rangle_k|-|\langle v_{dq}\rangle_k|\right)\\&\langle P_c \rangle_0=\sum\limits_{k=0,\pm 2}\left(\langle v_d\rangle_k\langle i_d\rangle_{-k}+\langle v_q\rangle_k\langle i_q\rangle_{-k}\right)\\
    &\langle \dot{\tilde{P}}_c \rangle_0 = \frac{1}{\tau_p}\left(\langle P_c \rangle_0-\langle \tilde{P_c} \rangle_0\right);~
    \langle \dot{\theta}_c \rangle_0=d_{pc}\left(\langle P_c^* \rangle_0-\langle \tilde{P_c} \rangle_0\right)
\end{aligned}
\end{equation}
\normalsize
Note that faster transients in $P_c (\tau)$ gets attenuated due to the low-pass filter and only $k = 0$ is considered for the dynamics of $\langle \tilde{P_c} \rangle$. As described in \eqref{eqn:dqtoDQ}, the same approximation is used for the dynamics of $\langle {\theta_c} \rangle$ .\\
2. \textit{Inner voltage controller (Fig.~\ref{fig:inner}(a)):} State-space equations of this controller can be expressed in DP form as
\small
\vspace{-3pt}
\begin{equation}
\begin{aligned}    &\langle\dot{x}_{1dq}\rangle_k=B_{11}\langle v_{dq}^*\rangle_k+B_{12}\langle v_{dq}\rangle_k-jk\omega_s\langle x_{1dq}\rangle_k\\
    &\langle i_{tdq}^*\rangle_k=C_{11}\langle x_{1dq}\rangle_k+D_{11}\langle v_{dq}^*\rangle_k+D_{12}\langle v_{dq}\rangle_k+D_{13}\langle i_{dq}\rangle_k
\end{aligned}
\end{equation}
\vspace{-3pt}
\normalsize
where, 
\small
\vspace{-3pt}
\begin{equation*}
\begin{aligned}
    &B_{11}=diag\left(1,1\right), ~B_{12}=-diag\left(1,1\right),~C_{11}=diag\left(k_{vi},k_{vi}\right)\\
    &D_{11}=diag\left(k_{vp},k_{vp}\right),D_{12}=\begin{bmatrix}-k_{vp} & -\omega_c C \\\omega_c C & -k_{vp} \end{bmatrix},D_{13} = diag(1,1)
\end{aligned}
\end{equation*}
\vspace{-3pt}
\normalsize

\begin{figure}
	\vspace{-5pt}
	\centering
\includegraphics[width= 0.45\textwidth]{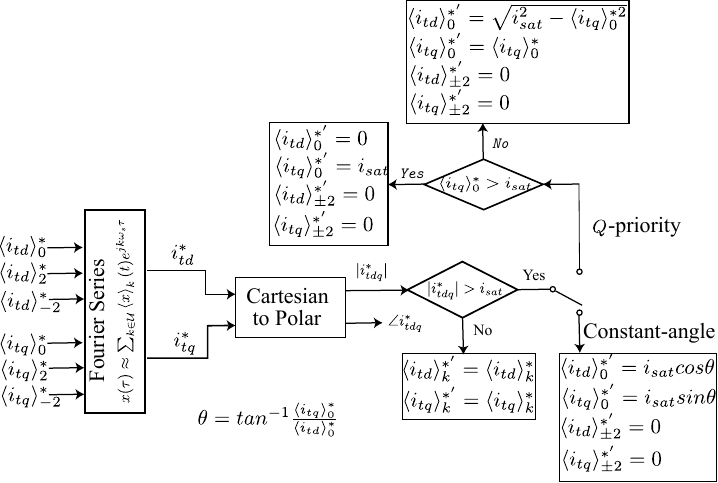}
	\caption{Current limiting strategy in DP framework.}
\label{fig:curlim}
	\vspace{-15pt}
\end{figure}

3. \textit{Current controller (Fig.~\ref{fig:inner}(a)):} Equation \eqref{eqn:curcontrol} explains the modeling of current controller in DP form.
\small
\vspace{-3pt}
\begin{equation}\label{eqn:curcontrol}
\begin{aligned}
&\langle\dot{x}_{2dq}\rangle_k=B_{21}\langle i_{tdq}^{*'}\rangle_k+B_{22}\langle i_{tdq}\rangle_k-jk\omega_s\langle x_{2dq}\rangle_k\\
    &\langle v_{tdq}\rangle_k=C_{21}\langle x_{2dq}\rangle_k+D_{21}\langle i_{tdq}^{*'}\rangle_k+D_{22}\langle i_{tdq}\rangle_k+D_{23}\langle v_{dq}\rangle_k
\end{aligned}
\end{equation}
\vspace{-3pt}
\normalsize
where,
\small
\vspace{-3pt}
\begin{equation*}
\begin{aligned}
    &B_{21}=diag\left(1,1\right), ~B_{22}=-diag\left(1,1\right),~C_{21}=diag\left(k_{ci},k_{ci}\right)\\
    &D_{21}=diag\left(k_{cp},k_{cp}\right),D_{22}=\begin{bmatrix}-k_{cp} & -\omega_c L \\\omega_c L & -k_{cp} \end{bmatrix},D_{23} = diag(1,1)
\end{aligned}
\end{equation*}
\normalsize
4. \textit{Current limiter (Fig.~\ref{fig:inner}(a)):} Between voltage controller and current controller, current limiting feature is incorporated to deal with overcurrent during short circuit fault. Depending on the system condition, either constant-angle or $Q$-priority current limiter will be activated. In practice, these strategies harness moving average operation to extract the dc component of the angle $\theta$ from $dq$ frame current quantity. Figure \ref{fig:curlim} demonstrates the current limiting features modeled in DP framework. Constant-angle current limiting strategy is mentioned below.  
\small
\begin{equation*}
    i_{td}^{*'} = \begin{cases}i_{td}^* & |i_{tdq}^*| \leqslant i_{sat}\\i_{sat}cos\theta & |i_{tdq}^*| > i_{sat}\end{cases}
    ;~i_{tq}^{*'} = \begin{cases}i_{tq}^* & |i_{tdq}^*| \leqslant i_{sat}\\i_{sat}sin\theta & |i_{tdq}^*| > i_{sat}\end{cases}
\end{equation*}
\normalsize
Here, $\theta=\tan^{-1}\left(\frac{\langle i_{tq}^*\rangle_0}{\langle i_{td}^*\rangle_0}\right)$ and $i_{sat}$ denotes $1.2$ times nominal current. The $Q$-priority current limiter is implemented as follows.
\small
\begin{equation*}
    i_{tq}^{*'} = \begin{cases}i_{tq}^{*} & |i_{tdq}^*| \leqslant i_{sat}\\\langle i_{tq}^*\rangle_0 & |i_{tdq}^*| > i_{sat}~\text{and}~\langle i_{tq}^*\rangle_0 < i_{sat}\\i_{sat} & |i_{tdq}^*| > i_{sat}~\text{and}~\langle i_{tq}^*\rangle_0 > i_{sat}\end{cases}
\end{equation*}
\vspace{-3pt}
\normalsize

\vspace{-15pt}
\small
\vspace{-3pt}
\begin{equation*}
    i_{td}^{*'} = \begin{cases}i_{td}^* & |i| \leqslant i_{sat}\\\sqrt{i_{sat}^2-\langle i_{tq}^*\rangle_0^2} & |i_{tdq}^*| > i_{sat}~\text{and}~\langle i_{tq}^*\rangle_0 < i_{sat}\\0 &|i_{tdq}^*| > i_{sat}~\text{and}~\langle i_{tq}^*\rangle_0 > i_{sat}\end{cases}
\end{equation*}
\vspace{-3pt}
\normalsize

5. \textit{Filter (Fig.~\ref{fig:inner}(b)):} The RLC filter in GFC is modeled in DP form using the following equation.
\small
\vspace{-3pt}
\begin{equation}
\begin{aligned}
   & \langle \dot{x}_{3dq}\rangle_k = A_{31}\langle x_{3dq}\rangle_k+B_{31}\langle v_{tdq}\rangle_k+B_{32}\langle v_{dq}\rangle_k-jk\omega_s\langle x_{3dq}\rangle_k\\
   & \langle i_{tdq}\rangle_k=C_{31}\langle x_{3dq}\rangle_k\\
    &\langle \dot{x}_{4dq}\rangle_k = A_{41}\langle x_{4dq}\rangle_k+B_{41}\langle i_{tdq}\rangle_k+B_{42}\langle i_{dq}\rangle_k-jk\omega_s\langle x_{4dq}\rangle_k\\
   & \langle v_{dq}\rangle_k=C_{41}\langle x_{4dq}\rangle_k
\end{aligned}
\end{equation}
\vspace{-3pt}
\normalsize
where,
\small

\begin{equation*}
\begin{aligned}
    &A_{31}=\begin{bmatrix}-\frac{R}{L} & \omega_c \\-\omega_c & -\frac{R}{L} \end{bmatrix},~A_{41}=\begin{bmatrix}0 & \omega_c \\-\omega_c & 0 \end{bmatrix}\\&B_{31}=B_{41}=diag\left(1,1\right),~B_{32}=B_{42}=-diag\left(1,1\right)\\ &C_{31}=diag\left(\frac{1}{L},\frac{1}{L}\right),
    ~C_{41}=diag\left(\frac{1}{C},\frac{1}{C}\right)
\end{aligned}
\end{equation*}
\normalsize
\subsection{Series-compensated transmission line model}
In order to model series-compensated transmission line in $pnz$ based DP, $k=\pm1$ Fourier coefficient is considered to approximate the actual signal. Based on the structure of our system, DP framework equations can be written as follows. Here, $R_L$ represents the load resistance that can be calculated from load real power $P_L$. Load reactive power $Q_L$ is assumed to be zero.
\small
\begin{equation}
   \begin{aligned}
       &L_{2pnz}\langle \dot{i}_{2pnz}\rangle_k=\langle v_{1pnz}\rangle_k-\langle v_{C2pnz} \rangle_k-\langle e_{bpnz}\rangle_k-\\&~~~~R_{2pnz}\langle i_{2pnz}\rangle_k-jk\omega_sL_{2pnz} \langle i_{2pnz}\rangle_k\\
      & L_{1pnz}\langle \dot{i}_{pnz}\rangle_k=\langle v_{pnz}\rangle_k-\langle v_{1pnz}\rangle_k-jk\omega_sL_{1pnz} \langle i_{pnz}\rangle_k\\
      & C_{2pnz}\langle \dot{v}_{C2pnz} \rangle_k = \langle i_{2pnz} \rangle_k-jk\omega_s C_{2pnz}\langle v_{C2pnz} \rangle_k \\
       &\langle v_{1pnz}\rangle_k = R_{Lpnz}\left(\langle i_{pnz}\rangle_k-\langle i_{2pnz}\rangle_k\right)
   \end{aligned} 
\end{equation}
\normalsize
where,
\small
\begin{equation*}
\begin{aligned}  L_{1pnz}&=diag\left(L_1,L_1,L_1\right),L_{2pnz}=diag\left(L_2,L_2,L_2\right), \\R_{2pnz}&=diag\left(R_2,R_2,R_2\right),C_{2pnz}=diag\left(C_2,C_2,C_2\right), \\R_{Lpnz}&=diag\left(R_L,R_L,R_L\right)
\end{aligned}
\end{equation*}
\normalsize
\subsection{Unbalanced fault modeling}
Unbalanced resistive fault is modeled by the bus fault impedance matrix in $pnz$ frame, which can be expressed as follows.
\small
\begin{equation*}
\begin{aligned}
    &R_{fabcg} = \begin{bmatrix} R_{fa}+R_g & R_g & R_g\\R_g & R_{fb}+R_g & R_g\\R_g & R_g & R_{fc}+R_g\end{bmatrix}\\
    &R_{fpnz} = T^{-1}R_{fabcg}T
\end{aligned}
\end{equation*}
\normalsize
where,
\small
\begin{equation*}
    T = \frac{1}{\sqrt{3}}\begin{bmatrix}
        1 & 1 & 1\\
        \alpha^2 & \alpha & 1\\
        \alpha & \alpha^2 & 1
    \end{bmatrix} \text{and} ~\alpha=e^{j\frac{2\pi}{3}}
\end{equation*}
\normalsize
Note that $R_{fa}$ and $R_g$ are phase $a$-to-neutral (similarly for phases $b$, $c$), and neutral-to-ground fault resistances, respectively. For example, phase $a$-to-ground fault can be simulated by setting $R_{fa}=R_f,~R_{fb}=\infty,~R_{fc} = \infty$, and $R_g=0$, where $R_f$ is the fault resistance. In practice, very large values are used for $R_{fb}$ and $R_{fc}$.
\small
\begin{table}[h!]
\caption{parameters of simulation model}
\vspace{-5pt}
    \label{tab:list_of_para}
    \begin{tabular}{c {l}c c}
         \hline
         \hline
         \textbf{Symbol} & \textbf{Description} & \textbf{Value} \\
         \hline
         $f_{s}$ & Nominal frequency ($Hz$) & $60$\\
         $R$ & AC side filter resistance ($m\Omega$) &  $0.722$\\
         $L$ & AC side filter inductance ($\mu H$) &  $44.4$\\
         $C$ & AC side filter capacitance ($F$) &  $0.0013$\\
         $L_1$ & Transformer leakage inductance ($mH$) &  $0.176$\\
         $R_2$ & Transmission line resistance ($\Omega$) &  $0.09$\\
         $L_2$ & Transmission line inductance ($H$) &  $0.0024$\\
         $C_2$ & Series compensation capacitance ($mF$) &  $3.59$\\
         $R_f$ & Fault resistance ($m\Omega$) & $0.756$\\
         $P_L$ & Load ($MW$) &  $100$\\
         $v$ & GFC terminal voltage ($kV$) &  $20.6$\\
         $e_b$ & Grid voltage ($kV$) &  $20$\\
         $P_c^*$ & Reference real power of GFC ($MW$) &  $400$\\
         $d_{pc}$ & Droop coefficient ($rad/s/MW$) &  $0.0174$\\
         $\tau_p$ & Droop controller delay ($ms$) & $10$\\
         $k_{p,ac}$ & Outer voltage controller proportional gain &  $0.001$\\
         $k_{i,ac}$ & Outer voltage controller integral gain ($s^{-1}$) &  $0.5$\\
         $k_{vp}$ & Inner voltage controller proportional gain ($\Omega^{-1}$) &  $2.34$\\
         $k_{vi}$ & Inner voltage controller integral gain ($\Omega^{-1}/s$) &  $5.22$\\
         $k_{cp}$ & Current controller proportional gain ($\Omega$) &  $0.16$\\
         $k_{ci}$ & Current controller integral gain ($\Omega/s$) &  $0.26$\\
         \hline \vspace{-5pt}
    \end{tabular}
\end{table}
\normalsize
\begin{figure}[!b]
	\vspace{-5pt}
	\centering
	\includegraphics[trim = {4cm 9.7cm 4cm 8.5cm}, clip,width= 0.35\textwidth]{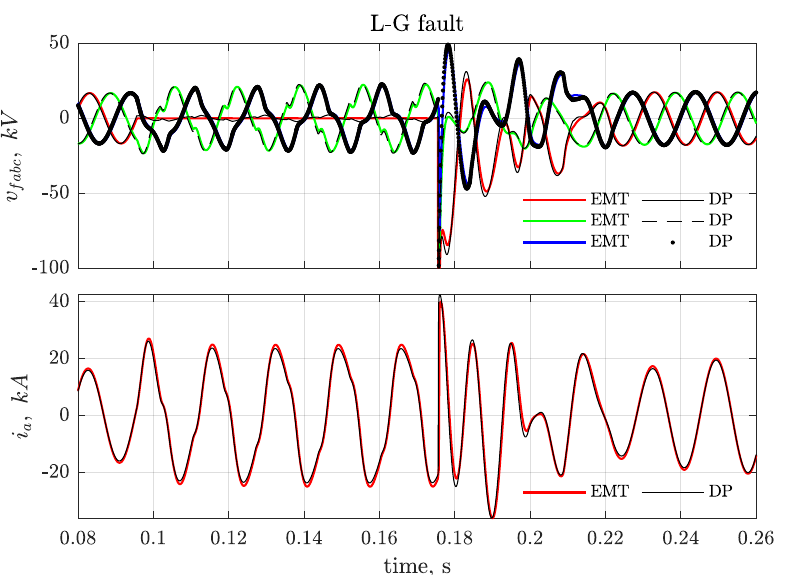}
	 \vspace{-10pt}
	\caption{Comparison of dynamic responses of EMTDC/PSCAD and DP model following a $L$-$G$ fault.}
	\label{fig:LG}
	\vspace{-15pt}
\end{figure}	

\begin{figure}
	\vspace{-5pt}
	\centering
	\includegraphics[trim = {4cm 9.7cm 4cm 8.5cm}, clip,width= 0.35\textwidth]{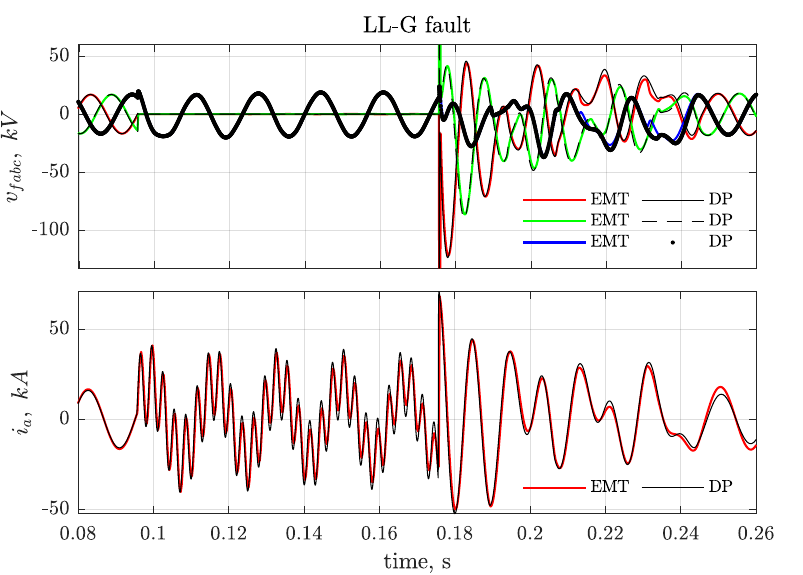}
	 \vspace{-10pt}
	\caption{Comparison of dynamic responses of EMTDC/PSCAD and DP model following a $LL$-$G$ fault.}
	\label{fig:LLG}
	\vspace{-15pt}
\end{figure}

\section{Results and Discussions}

Table I shows the parameters of the study system shown in Fig.~\ref{fig:TestSys}. 
Two cases for different compensation levels are considered in the DP model in Matlab/Simulink that is run with the variable-step solver $ode23tb$ \cite{simulink}. Results are compared with the EMT model which is constructed using the same parameters as the DP model adopted from  \cite{Lilan} in EMTDC/PSCAD \cite{pscad} using the averaged model of the GFC, which is run with a $1$ $\mu$s time step. Throughout the simulations, faults were created at $t=0.1$ s and cleared at $t=0.18$ s.

\subsection{Unbalanced fault simulation without GFC droop control}
\begin{figure}
	\vspace{-5pt}
	\centering
	\includegraphics[trim = {4cm 9.5cm 4cm 8.5cm}, clip,width= 0.35\textwidth]{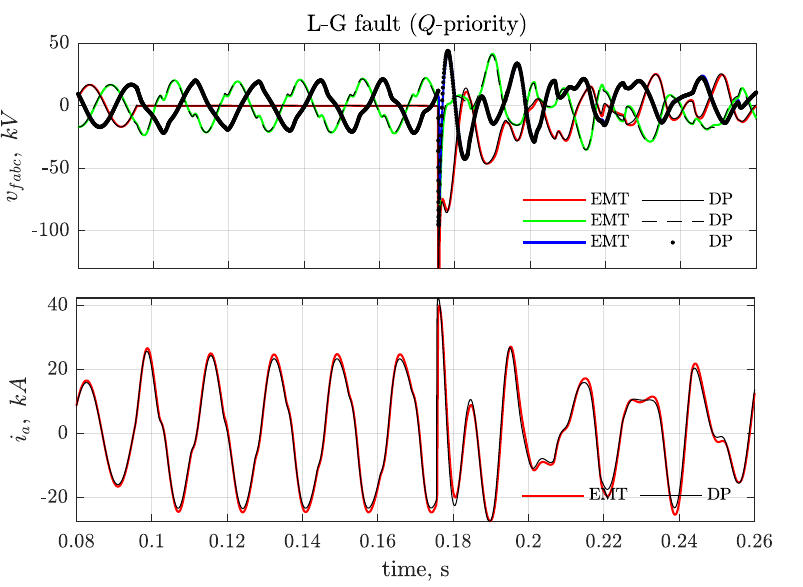}
	 \vspace{-10pt}
	\caption{Comparison of dynamic responses of EMTDC/PSCAD and DP model following a $L$-$G$ fault ($Q$-priority).}
	\label{fig:LGQ}
	\vspace{-15pt}
\end{figure}

\begin{figure}
	\vspace{-5pt}
	\centering
	\includegraphics[trim = {4cm 9.5cm 4cm 8cm}, clip,width= 0.35\textwidth]{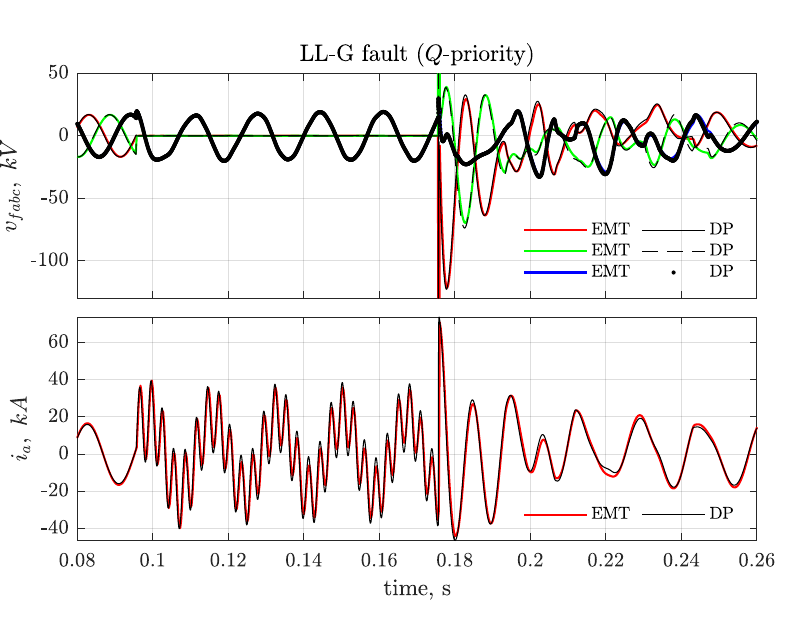}
	 \vspace{-10pt}
	\caption{Comparison of dynamic responses of EMTDC/PSCAD and DP model following a $LL$-$G$ fault ($Q$-priority).}
	\label{fig:LLGQ}
	\vspace{-20pt}
\end{figure}

Three types of faults, $L$-$G$, $LL$-$G$, and $LLL$-$G$ are applied at the load bus in Fig.~\ref{fig:TestSys} under $82$\% series compensation level where the GFC is operating without the droop control in Fig.~\ref{fig:Outer}(a). Due to space restriction, we show only a subset of unbalanced fault results. From Figs~\ref{fig:LG}, \ref{fig:LLG}, \ref{fig:LGQ}, and \ref{fig:LLGQ}, it is evident that the DP model provides near identical transient response like the EMT model for two different current limiting strategies. In case of $LL$-$G$ fault and $LLL$-$G$ fault (not shown), a higher frequency oscillation is observed that was associated with a $333$-Hz mode from the linearized DP model. Participation factor analysis shows that the mode originated from $C$, $L_2$, and $C_2$ in Fig.~\ref{fig:TestSys}. 
\vspace{-5pt}
\subsection{Stabilization of poorly-damped mode with higher series compensation in presence of GFC droop control}

\begin{figure}
	\vspace{-5pt}
	\centering
	\includegraphics[trim = {4cm 9.8cm 4cm 8.5cm}, clip,width= 0.4\textwidth]{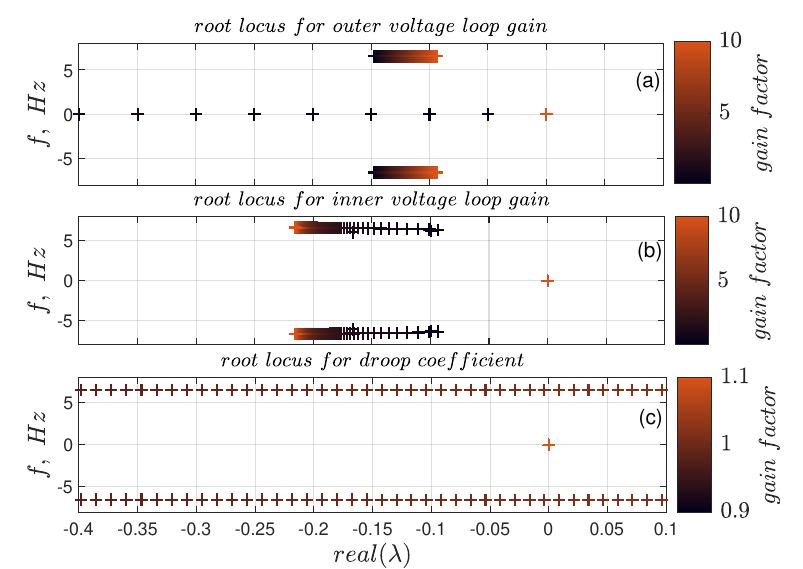}
	 \vspace{-10pt}
	\caption{Eigenvalue sensitivity with respect to (a) outer voltage loop gain, (b) inner voltage loop gain, and (c) droop coefficient.}
	\label{fig:EigSvty}
	\vspace{-8pt}
\end{figure}

\begin{figure}
	\vspace{-5pt}
	\centering
	\includegraphics[width= 0.5\textwidth]{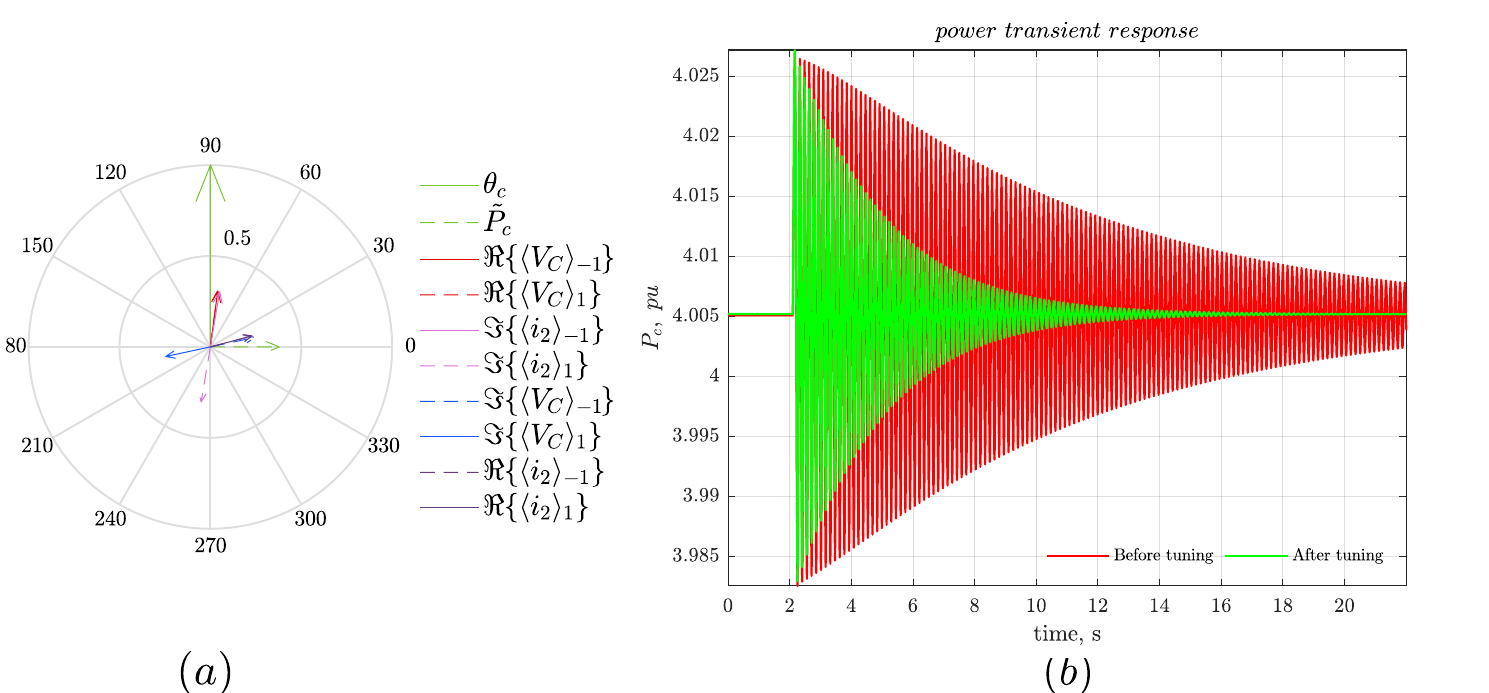}
	 \vspace{-10pt}
	\caption{(a) Compass plots of normalized participation factor magnitudes and modeshape angles of the dominant states contributing to the poorly-damped 6.54 Hz mode under 83.25\% series compensation. (b) EMT model's transient response due to pulse change in the reference of the outer voltage controller in Fig.~\ref{fig:Outer}(b).}
	\label{fig:compass_damp}
	\vspace{-20pt}
\end{figure}

When the GFC droop control is active, for a $83.25$\% compensation level, the linearized DP model shows that a $6.54$-Hz mode becomes very poorly damped. According to the compass plot in Fig.~\ref{fig:compass_damp}(a), the  voltage angle state $\theta_c$ has dominant participation in this mode. From this, we can interpret that the poor damping happens because the GFC is practically connected to a strong grid. Eigenvalue sensitivities with respect to outer voltage loop gains, inner voltage loop gains, and droop constant are analyzed by multiplying these by gain factors (gain factor is 1 for nominal values), which can be seen from Fig. \ref{fig:EigSvty}. It is evident from the root locus plots that this marginally stable mode is the most sensitive to the droop coefficient $d_{pc}$. Hence, $d_{pc}$ is slightly reduced by a factor $0.98$ to obtain a desired settling time of $15$ s. Note that reduced droop coefficient enables GFC to share more active power corresponding to a given change in frequency. As shown in Fig.~\ref{fig:compass_damp}(b), the dynamic response of the GFC power output in the EMT model following a pulse change in the reference of the outer voltage controller in Fig.~\ref{fig:Outer}(b) confirms the effectiveness of our proposed approach.


\section{Conclusions}
A linearizable DP-based modeling framework was proposed in this work that can seamlessly interface GFCs operated with conventional $dq$-frame based vector control with an unbalanced power systems. To that end, the GFC model was represented by $dq$-frame DPs of order $k = 0, \pm2$, which was interfaced with a series-compensated transmission line model in $pnz$-frame DPs of order $k = \pm1$. An approach for modeling two types of current limiting strategies, namely constant-angle and $Q$-priority in DP framework is proposed. Near-identical dynamic responses are observed from the DP-based model and the EMT-based model following unbalanced short circuit faults. Moreover, a poorly-damped oscillation beyond a certain level of series-compensation was observed that was primarily attributed to the GFC droop control based on the linearized DP model. Slight reduction of droop coefficient was needed to achieve a desired settling time that was verified using EMT simulation results. Our future research is directed towards scaling the DP model for multi-IBR bulk power systems.

\bibliographystyle{IEEEtran}
\bibliography{Mybib}

\end{document}